\documentclass[12pt]{article}
\usepackage[margin=1.2in]{geometry}
\usepackage{amsthm,amsmath,physics,mathtools,amssymb}

\usepackage{charter}

\usepackage{CJKutf8}

\usepackage{enumitem}

\theoremstyle{definition}

\usepackage{graphicx}
\graphicspath{ {images/} }

\usepackage{tensor}

\DeclarePairedDelimiterX{\inp}[2]{\langle}{\rangle}{#1, #2}

\usepackage{hyperref}
\hypersetup{
    colorlinks=true,
    linkcolor=blue,
    filecolor=magenta,      
    urlcolor=cyan,
}

\usepackage{cleveref}

\usepackage{xparse}

\NewDocumentCommand\LH{mo}{%
  \IfNoValueTF{#2}
   {\mathcal{L}(\mathcal{H}^{#1})}
   {\mathcal{L}(\mathcal{H}^{#1},\mathcal{H}^{#2})}%
}

\newcommand\id{\leavevmode\hbox{\small1\kern-3.3pt\normalsize1}}

\allowdisplaybreaks

\usepackage{authblk}

\usepackage[title]{appendix}

\usepackage{mciteplus}

\title{Time-space duality in 2D quantum gravity}

\author{Ding Jia (贾丁)\thanks{djia@perimeterinstitute.ca}}
\affil{Perimeter Institute for Theoretical Physics, Waterloo, Ontario, N2L 2Y5, Canada}
\affil{Department of Physics and Astronomy, University of Waterloo, Waterloo, Ontario, N2L 3G1, Canada}
\date{}

\begin{document}

\begin{CJK*}{UTF8}{gbsn}
\maketitle
\end{CJK*}

\begin{abstract}
An important task faced by all approaches of quantum gravity is to incorporate superpositions and quantify quantum uncertainties of spacetime causal relations. We address this task in 2D. 
By identifying a global $Z_2$ symmetry of 1+1D quantum gravity, we show that gravitational path integral configurations come in equal amplitude pairs with timelike and spacelike relations exchanged. 
As a consequence, any two points are equally probable to be timelike and spacelike separated in a universe without boundary conditions. In the context of simplicial quantum gravity we identify a local symmetry of the action which shows that even with boundary conditions causal uncertainties are generically present. Depending on the boundary conditions, causal uncertainties can still be large and even maximal.
\end{abstract}

\section{Introduction}\label{sec:i}


A great lesson of General Relativity is that spacetime is not a fixed background, but has its own dynamical laws. 
As shown by Hawking, King and McCarthy, and Malament \cite{hawking_new_1976, malament_class_1977}, up to a conformal factor a dynamical spacetime is completely determined by its causal relations. From this perspective, understanding spacetime is largely about understanding its dynamical causal structures.
In quantum theory, dynamical degrees of freedom are subject to quantum superpositions and exhibit quantum uncertainties. This poses two important questions to any approach to quantum gravity: How to incorporate quantum superpositions of spacetime causal relations? How large are the quantum uncertainties in spacetime causal relations?

The first question already has a solid answer.  Non-perturbatively defined gravitational path integrals generically sum over spacetime configurations of different causal structures. Therefore Lorentzian path integrals such as Lorentzian Quantum Regge Calculus \cite{Sorkin1974DevelopmentFields, Sorkin1975Time-evolutionCalculus, SorkinLorentzianVectors, Tate2011Fixed-topologyDomain, Tate2012Realizability1-simplex, AsanteEffectiveGravity},
(Locally) Causal Dynamical Triangulations \cite{Ambjorn2012NonperturbativeGravity, Jordan2013CausalFoliation, Jordan2013DeFoliation, Loll2015LocallyDimensions}, Lorentzian Spin-foam Models and Group Field Theories \cite{Perez2013TheGravity, Freidel2005GroupOverview}, Causal Sets \cite{Surya2019TheGravity} do incorporate quantum superpositions of spacetime causal relations.
The second question is harder. Quantifying quantum uncertainties in spacetime causal relations seems to require a non-perturbative calculation of Lorentzian gravitational path integrals. For this task, analytic solutions are hard to find, and numerical simulations face the sign problem. Even in lower dimensions where the theories are simpler, these challenges persist and the task appears daunting.

Nevertheless, we are able to provide an answer in 2D. Perhaps surprisingly, for a gravitational path integral without boundary conditions the quantum uncertainty in causal relations is in a certain sense maximal: Any pair of points are equally probable to be timelike and spacelike separated.

The answer is obtained, not by brute-force computations, but by recognizing symmetries. For a scalar field theory, the global $Z_2$ transformation acts as $\phi \mapsto -\phi$. We say the theory $Z_2$-symmetric if this leaves the action invariant. For gravity, the global $Z_2$ transformation acts as $g\mapsto -g$ on a metric field $g$. This does leave invariant the gravitational action 
\begin{align}
S=- \lambda \int d^2 x \sqrt{-g} + k \int d^2 x \sqrt{-g} R.
\end{align}
Since $\det g_{\mu\nu}=\det (-g_{\mu\nu})$ in 2D, the cosmological constant term is invariant. Since the Einstein-Hilbert term in 2D is topological \cite{Jee1984Gauss-BonnetSurfaces, Birman1984TheSpacetimes, Law1992NeutralManifolds, Chern1945OnManifold}, it is also invariant. In this sense, Lorentzian 2D quantum gravity has a global $Z_2$ symmetry.

\begin{figure}
    \centering
    \includegraphics[width=.5\textwidth]{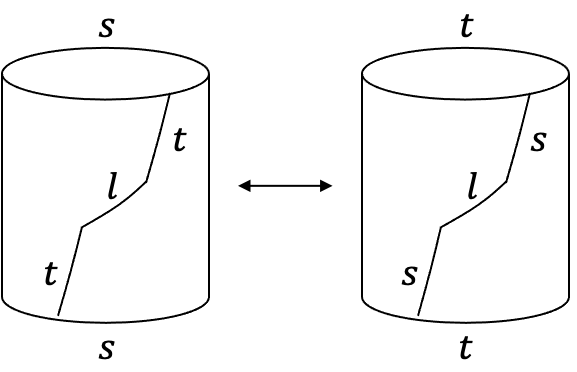}
    \caption{On a cylindrical surface, the time-space duality map leaves lightlike ($l$) line segments lightlike, but takes timelike ($t$) line segments to spacelike ($s$) line segments and \textit{vice versa}. The two configurations are physically inequivalent because one has spacelike boundaries and the other has timelike boundaries.}
    \label{fig:tsdm}
\end{figure}

Physically, the map $f(g):=-g$ exchanges timelike and spacelike separations. Under $f$, the line element $ds^2=g_{\mu\nu}dx^\mu dx^\nu$ changes sign, so a path will have its timelike pieces changed into spacelike pieces and \textit{vice versa} (\Cref{fig:tsdm}). Since the map $f(g):=-g$ is involutive, we will refer to it as the \textit{time-space duality map}. This map is not a diffeomorphism, because it generically relates physically distinct configurations (\Cref{fig:tsdm}).

Regarding causal relation uncertainties, whenever there is a configuration $g$ where a pair of points are timelike separated, there is the configuration $f(g)$ where they are spacelike separated. Since $g$ and $f(g)$ share the same action, they contribute equally to the path integral. Therefore the points are equally probable to be timelike and spacelike separated. This addresses the question of quantifying quantum uncertainties in spacetime causal relations with the path integral in the metric description without boundary conditions in 2D.

There are three questions left. What about theories with other variables? What about path integrals with boundary conditions? What about higher dimensions? These are the focus of the rest of this work. 
For 2D Locally Causal Dynamical Triangulations, it was already noted that with equal space and time distances for the elementary triangles, configurations related by an exchange of time and space contribute equally \cite{Loll2015LocallyDimensions}. The global $Z_2$ symmetry we point out here explains why this property is expected. In the following we will show that the global $Z_2$ symmetry and the conclusion of maximal causal uncertainty also holds in Lorentzian Quantum Regge Calculus. 
Working with Lorentzian Quantum Regge Calculus allows us to study the influence of the path integral measures, and incorporate topology changes and sum over topologies. In addition, in the presence of boundary conditions we will identify a local symmetry (which is not a gauge symmetry) of the action to show that although the causal uncertainty is no longer always maximal, it is generically present and can still be maximal for certain cases.
In higher dimensions, multiplying the metric by minus one does not yield a Lorentzian spacetime, and quantifying causal uncertainties in higher dimensions is left as an open question. However, the 2D results already have implications to understanding quantum gravity in higher dimensions, as we will discuss in the end.

\section{Lorentzian Quantum Regge Calculus}

\begin{figure}
    \centering
    \includegraphics[width=.65\textwidth]{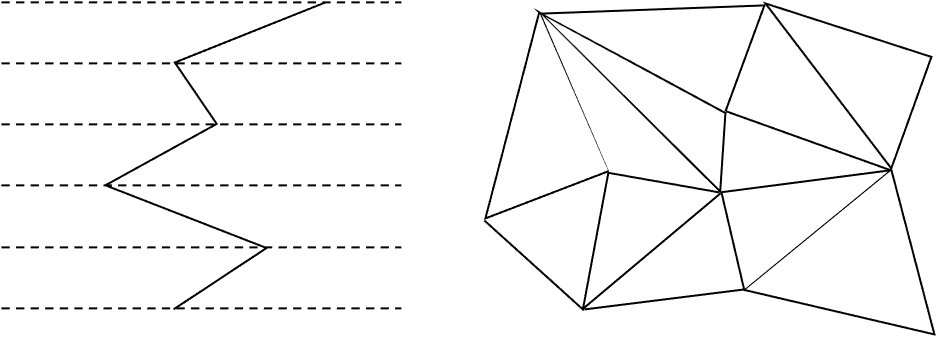}
    \caption{Piecewise linear paths and piecewise flat spacetimes.}
    \label{fig:plpf}
\end{figure}

In defining the path integral for a non-relativistic particle, we need to specify a way to enumerate the paths summed over. It is common to introduce to a temporal lattice, sum over piecewise linear paths on the lattice (\Cref{fig:plpf}), and take the continuum limit of lattice spaces going to zero.

In defining the path integral for gravity, we adopt a similar strategy to introduce a simplicial lattice, sum over piecewise flat spacetime geometries on the lattice (\Cref{fig:plpf}), and take the limit of moving to ever more refined lattices. This is of course the idea of Quantum Regge Calculus \cite{Rocek1981QuantumCalculus, Williams1992ReggeBibliography, Loll1998DiscreteDimensions, Barrett2019TullioGravity, Hamber2009QuantumApproach}, which builds upon Regge's insight \cite{Regge1961GeneralCoordinates} to use simplicial complexes to approximate curved spacetimes. While most previous works on Quantum Regge Calculus are for Euclidean spacetimes, here we work with Lorentzian spacetimes \cite{Williams1986QuantumFormulation, Sorkin1974DevelopmentFields, Sorkin1975Time-evolutionCalculus, SorkinLorentzianVectors, Tate2011Fixed-topologyDomain, Tate2012Realizability1-simplex, AsanteEffectiveGravity}.\footnote{As shown in \Cref{sec:as}, the ``spikes'' that cause troubles for certain Euclidean Quantum Regge Calculus models are eliminated by causal structure considerations in Lorentzian Quantum Regge Calculus.} 

In a Lorentzian 2D theory, we start with a simplicial lattice consisting of combinatorial triangles glued along their edges. The dynamical variables are the signed squared invariant distances assigned to the edges. These variables are the lattice analogs of the integrated line element $\int ds^2$ in the continuum, and we denote them by $\sigma_e$ where $e$ labels the edges. 
The relations $\sigma_e<,=,>0$ correspond to timelike, lightlike, and spacelike separations for the vertices of edge $e$. A spacetime configuration $\sigma=\{\sigma_e\}_e$ assigns a $\sigma_e$ value to every edge. This represents a piecewise flat spacetime geometry where all triangles are flat, but curvature need not be zero since the angles around a vertex need not sum to the flat spacetime value.

The partition function on a fixed simplicial lattice $\Gamma$ takes the form
\begin{align}\label{eq:pf}
Z_\Gamma =& \int \mathcal{D}[\sigma]~ C[\sigma] ~ e^{i(\lambda \sum_{t} A_t - 2\pi i \chi_\Gamma)}.
\end{align}
Due to a Lorentzian Gauss-Bonnet theorem \cite{SorkinLorentzianVectors}, the Einstein-Hilbert term $- 2\pi i \chi_\Gamma$ (which includes the boundary term contribution) is a topological invariant. The Euler number $\chi_\Gamma:=V-E+F$ depends only on the simplicial lattice $\Gamma$ through its vertices, edges, and faces numbers $V,E,F$.\footnote{See \Cref{sec:eht} for a review of the proof. 
} 
The cosmological constant ($\lambda$) term contains a sum over the areas $A_t$ of all triangles $t$. 

The Lorentzian flat triangle area formula \cite{AsanteEffectiveGravity, Tate2011Fixed-topologyDomain}
\begin{align}\label{eq:at}
A_t(a,b,c) = \frac{1}{4}\sqrt{a^2+b^2+c^2-2ab-2bc-2ac}
\end{align}
for edge $\sigma$ values $a,b,c$ can be obtained by embedding the triangle in Minkowski spacetime and evaluating the area there. 
Equation (\ref{eq:at}) is the same as Heron's formula for Euclidean triangles, except that the square root input has an extra global minus sign. For an Euclidean triangle, $A_t$ would therefore be purely imaginary. The distances $a,b,c$ belong to a Lorentzian triangle if and only if (\ref{eq:at}) is real. Spacetime configurations should only contain Lorentzian triangles. We enforce this in (\ref{eq:pf}) by the constraint $C[\sigma]$, which equals $1$ when $A_t\in\mathbb{R}$ for all triangles and equals $0$ otherwise. 

The path integral measure
\begin{align}\label{eq:mf}
\int \mathcal{D}[\sigma] =& \prod_e \int_{-\infty}^\infty d \sigma_e ~ \mu[\sigma]
\end{align}
contains a measure factor $\mu[\sigma]$ whose form has not been uniquely fixed in Quantum Regge Calculus (see Hamber \cite{Hamber2009QuantumApproach} Sections 2.4 and 6.9). One perspective is to view the measure ambiguity as similar to the action ambiguity in the Wilsonian approach to QFT. The task is then to identify universality classes within which the precise form of $\mu[\sigma]$ is less unimportant. We consider measures of the general form
\begin{align}\label{eq:mf2}
\mu[\sigma]=m[A_t,\abs{\sigma_e}] \prod_e \sigma_e^\alpha, 
\end{align}
where $m$ is an arbitrary function of the triangle areas $A_t$ and the unsigned squared invariant distances $\abs{\sigma_e}$, and $\alpha$ is a constant parameter.
This incorporates most, if not all measures studied previously in the literature \cite{Williams1992ReggeBibliography, Loll1998DiscreteDimensions, Barrett2019TullioGravity, Hamber2009QuantumApproach}, including the non-local ones discussed in \cite{Ambjrn1997SpikesCalculusb, Hamber1999OnGravity}.

To complete the definition of the path integral, we supply $Z_\Gamma$ of (\ref{eq:pf}) to the total partition function
\begin{align}\label{eq:pft}
Z =& \sum_\tau ~ \lim_{\Gamma\in \tau} Z_\Gamma,
\end{align}
which sums over a class of spacetime topologies after the lattice refinement limit is taken within the same topology $\tau$. Precisely which topologies should be included in the path integral is an open question of quantum gravity. The results in this work are independent of the answer to this question, so we will keep the sum $\sum_\tau$ flexible. 

Besides summing over spacetimes with different topologies, there is a separate concept of topology change within an individual spacetime \cite{SorkinLorentzianVectors, AsanteEffectiveGravity}. Without topology change, the path integral configurations are restricted so that each vertex has two lightcones and four light rays, and this condition can be put in $C[\sigma]$. When topology change is allowed, the vertices can have fewer or more than four light rays. 
The results in this work are independent of whether and what topology changes are allowed because the map $f$ leaves light rays as light rays, so we will keep these points flexible. 

\section{Global symmetry}


For simplicial geometries, the map $f$ multiplies $\sigma$ on all edges by $-1$:
\begin{align}
f: \{\sigma_e\}_e \mapsto \{-\sigma_e\}_e.
\end{align}
This leaves the triangle areas invariant since by (\ref{eq:at}), $A_t(a,b,c)=A_t(-a,-b,-c)$. Therefore the cosmological constant term of the action $\lambda\sum_t A_t$ is invariant. Since the topological Einstein-Hilbert term is also invariant, the whole action is invariant under $f$.

What about the constraint $C[\sigma]$ and the measure $\mu[\sigma]$? The constraint is invariant, i.e., $C[\sigma]=C[f(\sigma)]$, because as noted above $f$ preserves $A_t$. 
For the measure, $f$ takes (\ref{eq:mf2}) to
\begin{align}
\mu[-\sigma]=&m[A_t, \abs{\sigma_e}]\prod_e (-\sigma_e)^\alpha
\nonumber
\\
=& (-1)^{\alpha E} m[A_t, \abs{\sigma_e}] \prod_e \sigma_e^\alpha,
\end{align}
where $E$ is the total number of edges. If one forbids global complex phases to arise in $(-1)^{\alpha E}$ since they would depend on the artificial choice of lattice edge number $E$, then only integer $\alpha$ should be allowed. If $\alpha$ is odd, the path integral runs the risk of being identically zero. This is because for lattices with odd $E$, $\mu[\sigma]e^{iS[\sigma]}=-\mu[-\sigma]e^{iS[-\sigma]}$, so all pairs of configurations $(\sigma,-\sigma)$ cancel out in the path integral to make it identically zero. Therefore an odd $\alpha$ is unacceptable. The only option left is for $\alpha$ to be even. In this case $\prod_e \sigma_e^\alpha$ equals $\prod_e \abs{\sigma_e}^\alpha$ and can be absorbed into $m$ to make $\mu[\sigma]$ invariant under $f$. 

Altogether, for reasonable choices of $\alpha$ the general family of measure in (\ref{eq:mf2}) is invariant under $f$. The map $f$ is a $Z_2$ global symmetry of the theory without quantum anomalies.

\section{Maximal causal uncertainty}

We have shown that 1+1D gravitational path integral configurations come in pairs with equal amplitudes and with timelike and spacelike relations interchanged. Now we show that as a consequence, any two points are equally probable to be timelike and spacelike separated in an otherwise unconstrained path integral.

Consider any fixed configuration $\sigma$. The causal relation of any two vertices $(v_1,v_2)$ is determined by identifying a path with the smallest integrated unsigned invariant distance, and reading the causal signature of the path. Note that the path does not have to be a lattice path, but can cross into the simplices. For the partner configuration $f(\sigma)$, since $f$ preserves the integrated \textit{unsigned} invariant distances along all paths, the previous path stays as a path with the smallest integrated unsigned invariant distance, but with the opposite causal signature. This holds for both lattice paths and paths that cross into the simplices, because the interior of a simplex is equipped with the flat metric, which only changes sign under $f$. Therefore if $(v_1,v_2)$ are timelike (spacelike) separated in $\sigma$, they are spacelike (timelike) separated in  $f(\sigma)$.\footnote{When there are multiple extremal paths with opposite causal signatures in the original configuration, the dual configuration changes the causal signature of each path.}

Let $T$ and $S$ be the sets of configurations where $(v_1,v_2)$ are timelike and spacelike separated. By the previous analysis,
\begin{align}\label{eq:gpi}
    Z_\Gamma(T):= \int_{T} \mathcal{D}[\sigma] ~ C[\sigma] ~ e^{i(\lambda \sum_{t} A_t - 2\pi i \chi_\Gamma)} \nonumber \\
    =Z_\Gamma(S):= \int_{S} \mathcal{D}[\sigma] ~ C[\sigma] ~ e^{i(\lambda \sum_{t} A_t - 2\pi i \chi_\Gamma)}.
\end{align}
This implies $Z(T)=Z(S)$ for $Z$ in (\ref{eq:pft}) that can incorporate sums over topologies and topology changes.
Therefore the probability ratio for $(v_1,v_2)$ to be timelike separated and spacelike separated is $\abs{Z(T)}^2/\abs{Z(S)}^2=1$.\footnote{See \Cref{sec:ev} for an elaboration on measurement events.} Another implication is that the expectation value $\langle \sigma_e \rangle$ equals $0$ for any edge $e$, because positive and negative $\sigma_e$ values cancel in pairs.

\section{Local symmetry}


In the presence of boundary conditions (including those characterizing measurements on local boundaries) on a set of edges $\mathcal{B}$, the path integral
\begin{align}
Z_\Gamma(\alpha)=& \int \mathcal{D}[\sigma] ~ C[\sigma] ~ e^{i(\lambda \sum_{t} A_t - 2\pi i \chi_\Gamma)} \alpha(\{\sigma_e\}_{e\in \mathcal{B}})
\end{align}
is modified by a weight function $\alpha(\{\sigma_e\}_{e\in \mathcal{B}}) \in \mathbb{C}$ that describes the boundary conditions. For instance, in describing the result of a sharp measurement, $\alpha$ equals $1$ and $0$ respectively for configurations compatible and incompatible with that result. The global time-space symmetry no longer applies if the partner configurations have different $\alpha$. How does this affect causal uncertainties? 


It turns out that causal uncertainties are hard to constrain. Even when just one edge is subject to path integration and all other edges are constrained to take fixed values, that single edge could still exhibit large and even maximal causal uncertainty. Once more edges are unconstrained, their causal relations could also be very uncertain. 
We will now identify an edgewise local symmetry (which is interesting for its own sake) of the action to demonstrate these points.


Since the Einstein-Hilbert term (including possible boundary contributions) is topological (\Cref{sec:eht}), a local transformation is a symmetry transformation if it leaves the cosmological constant term invariant. Finding local symmetry transformations therefore amounts to finding local transformations that preserve the total spacetime area. We consider bulk and boundary edge transformations in turn. A transformation $h: \sigma_e\mapsto \sigma'_e$ acting on a bulk edge $e$ only affects the area of the two triangles $t_1$ and $t_2$ that contain $e$. Suppose the original total area of the two triangles is $A/4\in \mathbb{R}$, where the factor $1/4$ is brought in to cancel the same factor in (\ref{eq:at}). Then both $x=\sigma_e$ and $x=\sigma'_e$ should solve the same equation
\begin{align}\label{eq:cae}
A/4=A_{t_1}(x,a,b)+A_{t_2}(x,c,d),
\end{align} 
where $a,b,c,d$ are fixed $\sigma$ values on the other edges of the triangles. By (\ref{eq:at}) this equation can be used to derive a quadratic equation whose solutions are
\begin{align}\label{eq:qs}
x_\pm=&\frac{A^2 \left(u_1+u_2\right)-\left(u_1-u_2\right) \left(v_1-v_2\right)\pm 2A B}{4A^2-(u_1-u_2)^2},
\nonumber
\\
B=&\sqrt{A^4+A^2 \left(u_1 u_2-2 v_1-2 v_2\right)+(u_1-u_2)(u_1 v_2-u_2 v_1)+(v_1-v_2)^2},
\end{align}
where $u_1=2(a+b), v_1=(a-b)^2$, $u_2=2(c+d)$,  and $v_2=(c-d)^2$. When both $x_+$ and $x_-$ solve \eqref{eq:cae}, we define the local symmetry transformation $h$ on edge $e$ by
\begin{align}
h: \sigma_e\mapsto \sigma'_e,
\end{align}
where $\sigma_e$ is given as one of the solutions in (\ref{eq:qs}), and $\sigma'_e$ is the other solution.

For a boundary edge $e$, the difference is that $e$ is contained in just one triangle. The equation $A/4=A_t(x,a,b)$ is solved by
\begin{align}\label{eq:qs2}
x_\pm = a+b \pm\sqrt{A^2+4 ab}.
\end{align}
When both $x_+$ and $x_-$ solve \eqref{eq:cae}, the boundary local symmetry transformation $h$ is again defined to map one solution to the other.

By definition, the bulk and boundary local transformations leave the action invariant. The constraint $C[\sigma]$ enforcing $A_t\in \mathbb{R}$ for all triangles is also automatically fulfilled by the new configuration, because if not, the LHS of (\ref{eq:cae}) would have an imaginary part contradicting the assumption that $\sigma'_e$ solves (\ref{eq:cae}) with $A\in \mathbb{R}$. As explained below (\ref{eq:at}), the constraint $C[\sigma]$ ensures that the triangles are Lorentzian. Since the local transformation always fulfills the constraint $C[\sigma]$, one could not generate Euclidean triangles by applying the local transformation to Lorentzian triangles. However, the measure $\mu[\sigma]$ is not invariant in general, since $\abs{h(\sigma_e)}\ne \abs{\sigma_e}$ in general. Nevertheless, the often used family of measures $\mu[\sigma]=m[A_t]$ is invariant under $h$. 

Let us consider the implications on causal uncertainties when only one edge is subject to path integration. Since the causal signature of $\sigma_e$ is determined by its sign, $\sigma_e$ and $h(\sigma_e)$ have the same causal signature if their signs agree, and have the opposite causal signatures if their signs disagree. Therefore the range of $\sigma_e$ where $\sigma_e  h(\sigma_e) = x_+ x_-$ is positive (negative) is the range of causal (un)certainty. 
For (\ref{eq:qs}) and (\ref{eq:qs2}) respectively,
\begin{align}\label{eq:xpm1}
x_+ x_- =& \frac{A^4-2 A^2 \left(v_1+v_2\right)+\left(v_1-v_2\right){}^2}{-4 A^2+\left(u_1-u_2\right){}^2},
\\
\label{eq:xpm2}
x_+ x_- =& (a-b)^2-A^2 .
\end{align}
For both (\ref{eq:xpm1}) and (\ref{eq:xpm2}), $x_+ x_-$ always becomes negative for sufficiently large $A$. By inspecting (\ref{eq:at}) one sees that $A$ can always be made sufficiently large by increasing $x$, so the range of causal uncertainty is never empty. In this sense, causal uncertainty is still generically present in the presence of boundary conditions. On the other hand, the range of causal certainty can be empty, for instance when $v_1=u_1-u_2=0$. When this happens and when the measure is invariant under $h$, the causal uncertainty on $e$ is maximal, because the path integral configurations come in pairs of equal amplitude and opposite causal signatures on $e$.


When more than one edge is path integrated, the above reasoning can be applied to each unconstrained edge in turn. For the unconstrained edges, again causal uncertainties are generically present and could get large and even maximal in special cases. 

\section{Discussions}

We have identified two symmetries for quantum gravity in 1+1D. The global symmetry always interchanges timelike and spacelike relations, while the local symmetry  interchanges timelike and spacelike relations for some boundary conditions. These symmetries imply that in a universe without boundary conditions, any pair of points are equally likely to be timelike and spacelike separated. With boundary conditions, causal uncertainties are generically present, and can still be maximal in certain cases.








Gravity in 2D is special because the Einstein-Hilbert action is topological. It is worth discussing some questions raised by the 2D results that are relevant to higher dimensions.

Can large causal uncertainty signify the continuum limit for a 4D theory? Consider the probability ratio for some edge $e$ to be timelike separated against spacelike separated. This quantity, call it $r_e$, is defined in arbitrary dimensions in Lorentzian Quantum Regge Calculus, and is close to $1$ if causal uncertainty is large on the edge. We showed that $r_e=1$ exactly in 2D (without boundary conditions). What would $r_e$ be in 4D as the continuum limit is approached? The answer should be obtainable by numerical computation.\footnote{See \cite{AlexandruComplexProblem, Berger2019ComplexPhysics, Gattringer2016ApproachesTheory} for some reviews on methods to overcome the sign problem.} 
We speculate that $r_e$ gets close to $1$ in 4D in the continuum limit, first because of dimensional reduction \cite{Carlip2017DimensionGravity}, and second because quantum fluctuations generically get large at short distance scales. 
If this possibility is realized, then $r_e$ could be used as an order paramter, or ``observable'' in studying the continuum limit of lattice quantum gravity. This would confirm the expectation that indefinite causal structures play a central role in quantum gravity \cite{hardy2005probability, hardy2007towards}.

What would large causal uncertainty imply about the microscopic structure of quantum spacetime? Large causal uncertainties could imply effective discreteness for quantum spacetime, or from another perspective, an information-theoretic UV cutoff \cite{kempf2009information}. In information-theoretic models, very large causal uncertainties imply vanishing quantum correlations \cite{jia2018reduction, jia2017quantum} and communication capacities \cite{Jia2019CausalCommunication}. Heuristically, causal uncertainties obstruct the attempts to place one system in relation to another to harvest spacelike/timelike correlations, and the leakage of correlations to the environment results in a reduction of correlations for the designated systems. In quantum gravity it is also possible that no quantum information can be shared at microscopic scales.

Should unitarity and microcausality be fundamental requirements in quantum gravity? In a non-perturbatively defined gravitational path integral, the spacetime configurations do not share the same time foliation\footnote{Causal dynamical triangulation models assume a preferred time foliation \cite{Ambjorn2012NonperturbativeGravity}. The hope is that physical results agree with models that do not assume a preferred time foliation \cite{Jordan2013CausalFoliation, Jordan2013DeFoliation}, but currently it is unclear exactly for which cases this is true \cite{Loll2015LocallyDimensions}.} or the same causal structure. It is not even clear how to state the requirements of unitarity and microcausality, because unitarity refers to time evolution, and microcausality refers to spacelike separation. Since the symmetry transformations found here exchange timelike and spacelike relations, it is unlikely that unitarity and microcausality can ever be stated at the fundamental non-perturbative level in 2D. If unitarity and microcausality cannot be required in 2D, it seems not reasonable to regard them as fundamental requirements in quantum gravity. 

This does not mean that unitarity and microcausality cannot hold approximately, and it does not mean that unitarity and microcausality can be violated arbitrarily. In higher dimensions where the action is not topological and admits stationary points, spacetime configurations that dominate the path integral can be picked out by particular boundary conditions. Such stationary points could establish a connection to the classical spacetimes described in General Relativity. Unitarity and microcausality can be stated with respect to these semiclassical solutions. Since path integral configurations far away from the stationary points make subdominant contributions, large deviations from unitarity and microcausality are constrained. Unitarity and microcausality may be approximately true with respect to the semi-classical solution if the time evolution is close to unitary, and if the field commutator is close to zero for spacelike separated points, which is expected to be the case when the points are very far away. 

In this sense, contingent on boundary conditions, unitarity and microcausality could be effective though not fundamental requirements in quantum gravity. Contrary to a widespread misunderstanding, the absence of exact unitarity under a global time evolution does not logically imply the non-preservation of probabilities for measurement outcomes. For instance, any completely positive trace-preserving map preserves probabilities even if it is not unitary \cite{nielsen2000quantum}. The lack of global unitarity is not an obstacle to establish the normalization of outcome probabilities in gravity-matter coupled path integrals with local measurements.


\section*{Acknowledgments}
I am very grateful to Seth Asante, Bianca Dittrich, and Lee Smolin for valuable discussions on Lorentzian Quantum Regge Calculus, and to Lucien Hardy, Achim Kempf, Robert Mann, and Laurent Freidel for valuable discussions on quantum gravity in general.

Research at Perimeter Institute is supported in part by the Government of Canada through the Department of Innovation, Science and Economic Development Canada and by the Province of Ontario through the Ministry of Economic Development, Job Creation and Trade. This publication was made possible through the support of the grant ``Operationalism, Agency, and Quantum Gravity'' from FQXi. The opinions expressed in this publication are those of the author and do not necessarily reflect the views of the funding agencies. 

\newpage

\appendix

\begin{appendices}
\section{Absence of spikes}\label{sec:as}

\begin{figure}
    \centering
    \includegraphics[width=.6\textwidth]{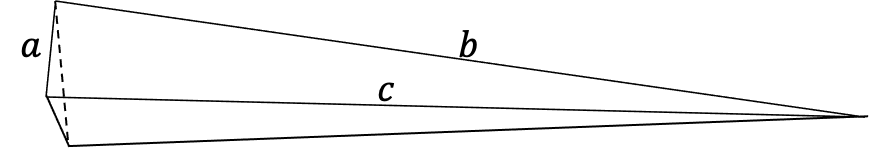}
    \caption{A spike consisting of three triangles glued along their long interior edges. The short edge with $\sigma$ value $a$ belongs to the perimeter.}
    \label{fig:spike}
\end{figure}

``Spikes'' are regions with a small total area, a small perimeter, but very long interior edges. Euclidean Quantum Regge Calculus admits spikes. The area for an Euclidean triangle with squared edge lengths $a,b,c$ is
\begin{align}\label{eq:ate}
A_t(a,b,c) =& \frac{1}{4}\sqrt{-a^2-b^2-c^2+2ab+2bc+2ac}
\\
=& \frac{1}{4}\sqrt{a(2b+2c-a)-(b-c)^2}.
\end{align}
When $b=c\gg 1$ and $a=1/b$, we have $A_t \sim 1$. A number of such triangles can be composed along their long edges to form a spike (\Cref{fig:spike}).

Euclidean Quantum Regge Calculus faces the challenge that for a family of measures, spikes cause diverging expectation values for powers of edge lengths. On any edge there exists a number $n$ so that the expectation value of edge length to the $n$-th power, $\langle l^n \rangle$, diverges even when the total spacetime area and the perimeter are bounded \cite{Ambjrn1997SpikesCalculusb}.

The case is different in the Lorentzian. 
The absence of spikes in the Lorentzian had been shown in \cite{Tate2011Fixed-topologyDomain} in the restricted setting with hexagonal lattice whose edges have certain fixed causal signatures. Here we show that spikes are absent in Lorentzian 2D Quantum Regge Calculus even without these restrictions. The only condition we need is that all vertices have at least one emanating light ray. In an ordinary 2D spacetime configuration, four light rays emanate from each point so that two light cones are formed. Exotic spacetime configuration can have vertices emanating a different number of light rays. We show that spikes can only exist in very exotic spacetime configurations: The tip of a spike can emanate no light rays, so there must be no lightcones at the tip of the spike. 
Consequently, demanding that every vertex has at least one emanating light ray rules out spikes.

The area for a Lorentzian triangle with signed squared edge distances $a,b,c$ is
\begin{align}\label{eq:at1}
A_t(a,b,c) =& \frac{1}{4}\sqrt{a^2+b^2+c^2-2ab-2bc-2ac}
\\
=& \frac{1}{4}\sqrt{-a(2b+2c-a)+(b-c)^2}.\label{eq:at2}
\end{align}
To have spikes, there should be triangles with two large $\abs{\sigma}$, one small $\abs{\sigma}$, and a small area. 

\begin{figure}
    \centering
    \includegraphics[width=.6\textwidth]{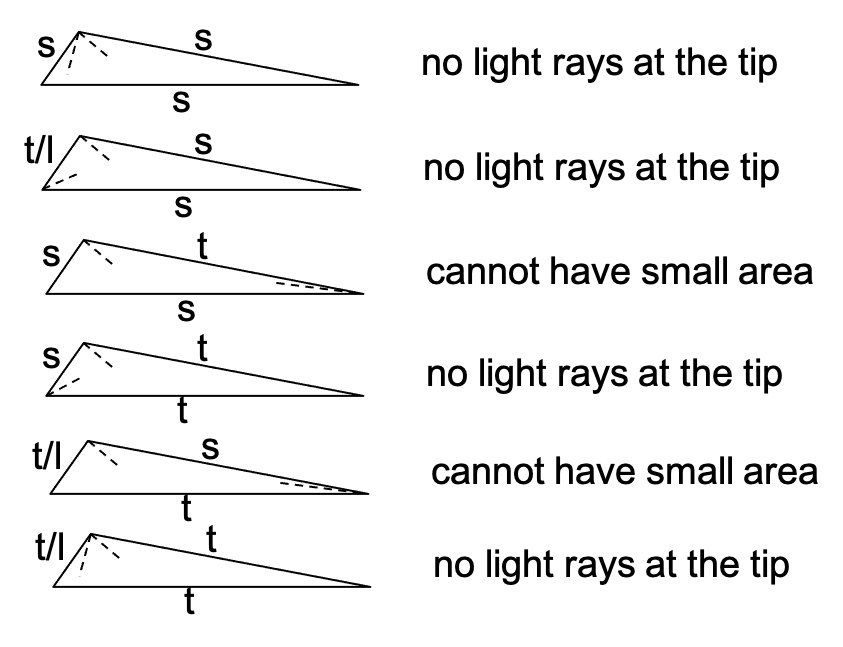}
    \caption{Triangles with one small $\abs{\sigma}$ (on the left edge) and two large $\abs{\sigma}$. The letters t, l, s stand for timelike, lightlike, and spacelike separations. Only the left edge can be lightlike because lightlike $\abs{\sigma}$ is zero and hence small. The number of dashed lines at a vertex represents the number of light rays at that vertex.}
    \label{fig:ns}
\end{figure} 

In \Cref{fig:ns} we list all triangles with two large $\abs{\sigma}$ and one small $\abs{\sigma}$ according to the causal signatures of the edges. We claim that they either cannot have any light rays at the tip, or cannot have a small area. Consequently the tip of a spike cannot emanate any light rays.

To establish the claim, note that the number of light rays at a vertex can be found by embedding the flat triangle in Lorentzian spacetime and counting the number of light rays there. It is not hard to show that in a triangle with both causal (timelike/lightlike) and acausal (spacelike) edges, vertices connected to both kinds of edges have one light ray, while other vertices have no light rays. In a triangle where all edges are causal or acausal, the vertex facing the longest edge has two light rays, while the other vertices have no light rays. These determine the dashed lines in \Cref{fig:ns}. Regarding area, the third triangle in \Cref{fig:ns} cannot have a small area, because without loss of generality we can let $a\ll -1$ represent the long timelike edge in (\ref{eq:at2}) and it becomes clear that $A_t\gg 1$. Similarly, the fifth triangle cannot have a small area, because without loss of generality we can let $a\gg 1$ represent the long spacelike edge in (\ref{eq:at2}) and it becomes clear that $A_t\gg 1$.

\section{The Einstein-Hilbert term}\label{sec:eht}

\begin{figure}
    \centering
    \includegraphics[width=.3\textwidth]{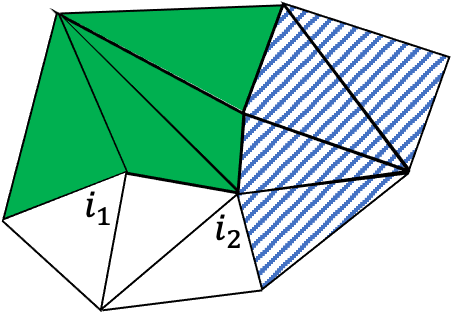}
    \caption{Boundary vertices shared by multiple regions. Three (shaded, striped, blank) regions are shown. Vertex $i_1$ belongs to two regions, while vertex $i_2$ belongs to three regions.}
    \label{fig:bv}
\end{figure}

Here we review a proof by Sorkin \cite{SorkinLorentzianVectors} that the Einstein-Hilbert action in Lorentzian 2D Regge Quantum Calculus is topological. The original proof assumes that each vertex belongs to exactly two bounded regions. More generally, a vertex can be shared by two or more bounded regions (\Cref{fig:bv}). The proof we show below holds for the general case. Since the global and local symmetry transformations studied in this work do interfere with this issue, all results still hold in this more general case.

In 2D Regge Calculus, we have the following correspondence between continnum and lattice expressions \cite{Hamber1986SimplicialDimensions}\footnote{See Section 6.5 of Hamber \cite{Hamber2009QuantumApproach} for an introduction to the Regge analogs of the gravitational action.}: 
\begin{align}
& R(x) \rightarrow \delta_v/A_v,\label{eq:eha1}
\\
&\sqrt{-g(x)} \rightarrow A_v,
\\
&\int d^2 x \sqrt{-g(x)} R(x) \rightarrow \sum_v \delta_v, \label{eq:eha}
\end{align}
where $\delta_v$ is the deficit angle at vertex $v$, and $A_v$ is the area at vertex $v$. There are different ways to obtain $A_v$ from the triangle areas $A_t$. However, these differences are irrelevant to the Einstein-Hilbert action (\ref{eq:eha}), which is simply the sum of the deficit angle over all vertices. 

In the Lorentzian, the deficit angle is \cite{SorkinLorentzianVectors}
\begin{align}\label{eq:bda}
\delta_v := -2\pi i - \sum_{j\in v} \theta_j,
\end{align}
where the sum is over all triangle angles $\theta_j$ tipped at the vertex $v$. This is called the ``deficit angle'', because it is the difference between the total angle around a vertex in Lorentzian flat spacetime, $-2\pi i$, and the total angle of the triangles around vertex $v$, $\sum_{j\in v} \theta_j$. The more the total angle $\sum_{j\in v} \theta_j$ differs from the flat spacetime value, the more spacetime curves, and the larger $\abs{\delta_v}$ gets. Indeed the relations (\ref{eq:eha1})-(\ref{eq:eha}) show how the deficit angle quantifies spacetime curvature.

When multiple bounded spacetime regions $\{R_k\}$ are composed along their boundaries, we want to maintain additivity for the action \cite{Hartle1981BoundaryCalculus, SorkinLorentzianVectors}:
\begin{align}
S(\cup_k R_k)=\sum_k S(R_k).
\end{align}
In the following we will set the constant prefactor of the action to $1$ for simplicity. Suppose a boundary vertex $v$ is shared by $n_v$ many bounded regions $\{R_k\}$ and becomes a bulk vertex after the composition. Then the boundary terms $S_v(R_k)$ at $v$ in regions $R_k$ should obey
\begin{align}
\sum_k S_v(R_k) = \delta_v = -2\pi i - \sum_{j\in \cup_k R_k} \theta_{j}.
\end{align}
One way to meet this requirement is to assign
\begin{align}\label{eq:ba}
S_v(R_k) = -\frac{2\pi i}{n_v} - \sum_{j\in R_k} \theta_{j}.
\end{align}
This boundary term at vertex $v$ depends on the number $n_v$ of regions that share $v$. 
This piece of information cannot be read from the spacetime configuration of region $R_k$, but must be supplied externally.

We are now ready to present the proof that the Lorentzian Einstein-Hilbet action is topological, i.e., the Lorentzian Gauss-Bonnet theorem, for arbitrary regions $R_k$ with or without boundary. The proposition is that
\begin{align}\label{eq:lgbt}
S(R_k) = -2\pi i \chi, \quad \chi= V^{\mathrm{o}} + \frac{1}{2} V^\partial + \sum_{v\in \partial} \frac{1}{n_v} - E + F,
\end{align}
where $V,E, F$ are the vertex, edge, and face numbers of the simplicial lattice, with the bulk and boundary contributions are labelled by superscripts $\mathrm{o}$ and $\partial$. The sum $\sum_{v\in \partial}$ is over all boundary vertices. In the special case that each boundary vertex is shared by two regions, i.e., $n_v=2$, $\chi$ reduces to the familiar Euler number $\chi=V-E+F$.

To prove the proposition, we use
\begin{align}\label{eq:gb1}
S/(-\pi i)=&2V^{\mathrm{o}} + \sum_{v\in \partial} \frac{2}{n_v} - F,
\\
0=&-2 E^{\mathrm{o}} - E^\partial + 3F,
\label{eq:gb2}
\\
0=&V^{\partial} - E^\partial.
\label{eq:gb3}
\end{align}
To obtain (\ref{eq:gb1}), we first sum over the bulk and boundary contributions to the Einstein-Hilbert action according to (\ref{eq:bda}) and (\ref{eq:ba}). The constant $-2\pi i$ terms sum to the first two terms on the RHS of (\ref{eq:gb1}). The $\theta_j$ terms sum over the all the triangle angles of the region. Since all triangle of the region are flat and their interior angles sum to $\sum_{j\in t}\theta_j = -\pi i$ \cite{SorkinLorentzianVectors}, this gives rise to the $F$ term of the RHS of (\ref{eq:gb1}). Equations (\ref{eq:gb2}) and (\ref{eq:gb3}) are simple facts about the simplicial lattices. Each bulk edge is shared by two faces, each boundary edge is shared by one face, and each face has three edges so (\ref{eq:gb2}) holds. The boundary is formed by a vertex-edge-vertex-edge... chain so (\ref{eq:gb3}) holds. Adding up (\ref{eq:gb1}) to (\ref{eq:gb3}) yields (\ref{eq:lgbt}).


\section{Events}\label{sec:ev}

Consider the path integral of a non-relativistic particle. Suppose the particle was seen at position $x$ at time $t_1$, and we want to calculate the probability ratio of seeing the particle at position $y$ and at position $z$ at time $t_2>t_1$. The relevant path integrals are
\begin{align}\label{eq:ppi1}
Z(y)=&\int_{x'(t_1)=x,x'(t_2)=y} \mathcal{D} x'~ e^{iS},
\\
Z(z)=&\int_{x'(t_1)=x,x'(t_2)=z} \mathcal{D} x'~ e^{iS},\label{eq:ppi2}
\end{align}
and the probability ratio is given by $r=\abs{Z(y)}^2/\abs{Z(z)}^2$.

Here the first set of events is ``seeing the particle at position $x$ at time $t_1$ and at position $y$ at time $t_2$'', while the second set of events is ``seeing the particle at position $x$ at time $t_1$ and at position $z$ at time $t_2$''. These sets of events translate into restrictions on the path integration: Integrate only over particle path configurations compatible with the events.

Strictly speaking the above path integral description of the events is incomplete. The event is not just that the particle reaches certain positions at certain times, but that some observer sees the particle reaching these positions at certain times. A complete description should include the observer's physical system, and the path integration should be over the joint particle-observer configurations compatible with the events. 

Such a complete description will yield a probability ratio $r'$ that can in principle differ from $r$ derived from (\ref{eq:ppi1}) and (\ref{eq:ppi2}). When the influence of the observer's system is weak, $r$ approximates $r'$ well, and (\ref{eq:ppi1}) and (\ref{eq:ppi2}) can be used in practice.

The case of quantum gravity is similar. Here the gravitational configurations are analogous to the particle path configurations. The probability ratio between timelike and spacelike separations derived from (\ref{eq:gpi}) 
is an approximation. It approximates the result from a more complete path integral containing both gravity and matter describing some observers detecting two ``spacetime events'' to be timelike and spacelike separated. Here a ``spacetime event'' corresponds to the physical constituents of some observer(s) and the material and gravitational surrounding taking some particular arrangements. The path integration is over matter and spacetime configurations compatible with such arrangements. 

The probability ratio from this more complete path integral containing both gravity and matter can in principle differ from that from (\ref{eq:gpi}), but when matter backreaction is weak, (\ref{eq:gpi}) can be used in practice.
\end{appendices}

\bibliographystyle{unsrt}
\bibliography{mendeley.bib}

\end{document}